\documentclass[checkin,noshowpacs,aps,prl,noshowkeys]{revtex4}
\usepackage{threeparttable}
\usepackage{amsmath}
\usepackage{multirow}
\usepackage{booktabs}
\usepackage{graphicx,color}
\usepackage{graphicx}
\usepackage{subfigure}
\usepackage{amsmath}
\newcommand{\ba}{\begin{eqnarray}}
\newcommand{\ea}{\end{eqnarray}}

\newcommand{\be}{\begin{equation}}
\newcommand{\ee}{\end{equation}}

\definecolor{pink}{rgb}{1,0.18,1.0}
\def\prl{{ Phys. Rev. Lett. }}

\def\prb{{ Phys. Rev. B }}

\def\jcp{{ J. Chem. Phys. }}

\def\sci{{ Science }}

\def\jap{{J. Appl. Phys. }}

\def\nl{{Nano Lett. }}
\def\nn{{Nat. Nanotech. }}
\def\acsnn{{ACS Nano }}

\def\nc{{Nat. Commun. }}
\def\nnt{{Nat. Nanotech. }}

\def\jpcl{{J. Phys. Chem. Lett. }}
\def\sci{{Science }}

\def\pr{{Phys. Rev. }}

\begin{document}

%\title{Unexpected semiconducting properties in few-layer arsenene}

\title{Manifestation of unexpected semiconducting properties in
  few-layer orthorhombic arsenene}

\author{Zhiya Zhang}
\affiliation{Key Laboratory for Magnetism and Magnetic Materials of
 the Ministry of Education, Lanzhou University, Lanzhou 730000, China}

\author{Jiafeng Xie}
\affiliation{Key Laboratory for Magnetism and Magnetic Materials of
 the Ministry of Education, Lanzhou University, Lanzhou 730000, China}

\author{Dezheng Yang}
\affiliation{Key Laboratory for Magnetism and Magnetic Materials of
 the Ministry of Education, Lanzhou University, Lanzhou 730000, China}

\author{Yuhua  Wang}
\affiliation{Key Laboratory for Magnetism and Magnetic Materials of
 the Ministry of Education, Lanzhou University, Lanzhou 730000, China}

\author{Mingsu Si$^{*}$}
\affiliation{Key Laboratory for Magnetism and Magnetic Materials of
 the Ministry of Education, Lanzhou University, Lanzhou 730000, China}

\author{Desheng Xue$^{\dagger}$}
\affiliation{Key Laboratory for Magnetism and Magnetic Materials of
 the Ministry of Education, Lanzhou University, Lanzhou 730000, China}

\date{\today}

\begin{abstract}
In this express, we demonstrate few-layer orthorhombic arsenene is an
ideal semiconductor. Due to the layer stacking, multilayer arsenenes
always behave as intrinsic direct bandgap semiconductors with gap
values of around 1 eV.  In addition, these bandgaps can be further
tuned in its nanoribbons.  Based on the so-called acoustic phonon
limited approach, the carrier mobilities are predicted to approach as
high as several thousand square centimeters per volt-second and 
simultaneously exhibit high directional anisotropy.  All these make few-layer arsenene
promising for device applications in semiconducting industry.  
\end{abstract}

%solid/electronic structure,fullerenes/electrical
%conductivity,/magnetic properties/monolayers 

%\pacs{73.20.-r, 73.61.Cw, 61.46.-w, 78.30.Am}

%\keywords{few-layer orthorhombic arsenene, proper bandgaps, high
%  carrier mobilities, first-principles calculations} 

\maketitle

%\section{I. Introduction} 

Mainstream logic devices applications  depend largely on
high performance of semiconducting materials \cite{nl_cui}. Graphene exhibiting
superior carrier mobility seems to be an ideal candidate. On the other
hand, the ability to control the electronic properties of a material by external
field requires a moderate electronic bandgap \cite{sci_nov}. Unfortunately, the
intrinsic dispersion of graphene is gapless. Entire community
still contributes to search for novel semiconducting materials.
Recently, phosphorene, monolayer of black phosphorus (BP), is emerging
as a promising two-dimensional (2D) semiconductor which might go
beyond graphene.
Phosphorene  holds a high carrier mobility in a wide range of
$\sim$10$^{3}$-10$^{4}$ cm$^{2}$V$^{-1}$s$^{-1}$ \cite{li_nn,ji}, and
simultaneously  maintains a considerable bandgap of around 2.0 eV
\cite{nl_liang}. At the same time, phosphorene and its bulk
counterpart BP are stimulating intense fundamental researches owing to
its versatile properties. For instance, structural transformations at high pressure
\cite{jamieson}, low-temperature superconducting \cite{wittig}, a
highly anisotropic transport \cite{liu_acs}, excellent optical and
thermoelectric responses \cite{nl_fei0}, a negative Poisson's ratio
\cite{jiang}, and strain-induced inversion of conduction bands
\cite{nl_fei} have been demonstrated. It is quite likely that these
unusual properties make phosphorene potential for realization of
optoelectronic devices in post-silicon era.

With five ``$sp$'' electrons in valence state, As sits below and
closest to P in the periodic table. The condensed state substances of
these two elements are most likely similar with each other from the
chemical point of view. However, up to now, little is known for As
compared to black phosphorene \cite{kamal,jianw}.
On the other hand, following the success of graphene in experiments,
various chemical classes of 2D materials initially considered
to exist only in the realm of theory have been synthesized. Taking
silicene as an example, it is first proposed through using   
{\it ab initio} method \cite{takeda,prl_cahangirov} and then epitaxially
grown on both silver (111) substrate \cite{prl_vogt} and diboride thin films
\cite{prl_fleurence} in experiments. All these  motivate a
timely study of As in low-dimensional forms, which would be
promising for future device applications.

This express reports a large bandgap and a high carrier mobility
simultaneously  appearing in few-layer orthorhombic arsenic, based on
first-principles  calculations. 
The chemical symbol of As is the abbreviation of the
word ``arsenic''. Thereafter, a monolayer of As can be called
``arsenene'', in analogy with graphene and phosphorene.
Although the bandgap of the monolayer structure was reported
to be indirect \cite{kamal}, our results surprisingly
show that multilayer arsenenes always behave as intrinsic direct
bandgap semiconductors with proper bandgaps around 1 eV. 
More importantly, the carrier mobilities are
predicted to be as high as several thousand square centimeters per
volt-second.  All these make few-layer arsenene intriguing for device 
applications in semiconducting industry.

\begin{figure}
\includegraphics[width=16cm]{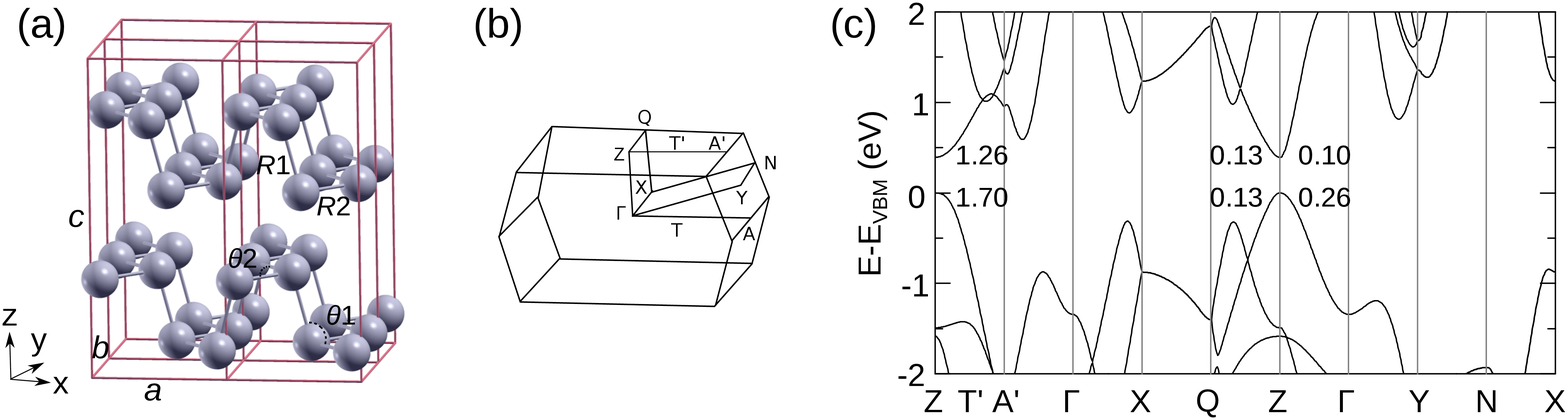}
\caption{(color online) (a) The conventional crystal structure of
orthorhombic bulk As. The lattice constants $a$, $b$, and $c$ and the
internal parameters $R$1, $R$2, $\theta$1, and $\theta$2 are given as
well. A 2$\times$2$\times$1 supercell is taken for clarity. (b) First
Brillouin zone of primitive unit cell. The high-symmetry lines and
points are given as well. 
(c) Band structure for orthorhombic As calculated using the HSE06 hybrid functional
under the revPBE-vdW functional optimized structure. The
fitted effective masses are given along the Z-T$^{'}$-A$^{'}$, Z-Q,
and Z-$\Gamma$ directions. E$_{\rm VBM}$ is the energy of valence-band maximum.}
\label{fig1}
\end{figure}

The most stable configuration of As allotropes is the rhombohedral
(A7) structure (called grey As). This is a metallic phase as bands near the T point and
at the L point largely overlap by about 0.5 eV \cite{xu}.  When it is
heated at the boiling point of water ($\sim$370 K), an orthorhombic
phase (arsenolamprite) arises as a similar structure to BP
\cite{krebs}. Owing to the less dense density
compared with the A7 phase, the orthorhombic As is a narrow-band
semiconductor with bandgap in the order of 0.3 eV \cite{greaves}.
 In reality, little attention has been paid to the
crystalline As in particular the orthorhombic one. The much
sparser amorphous As with a gap of 1 eV or more
takes up most of the research interest \cite{pollard}.
  In this context, we focus our study on the orthorhombic As, as shown
  in Fig. \ref{fig1}(a). First Brillouin zone of the primitive cell is
  also displayed  in  Fig. \ref{fig1}(b).
To guarantee the lattice parameters used accurately, five functionals are taken in
realistic simulations (see Methods in Supporting Information (SI)) \cite{si}. The
optimized structural parameters are summarized in 
Tab. I. In comparison with the experiment \cite{smith}, the revPBE-vdW
functional \cite{rev} yields the
reasonable lattice parameters which are slightly larger than the
experimental ones. By contrast, the functional of PBE
\cite{pbe} without vdW correction produces a large
error along the layer-stacking direction by as large as 4.1 percent. This 
means the vdW correction is of crucial importance in determining the
interlayer distance.  Generally, pure DFT underestimates the bandgap of
a semiconductor because of the improper consideration of Coulomb
interaction. The
functional of revPBE-vdW gives a near zero bandgap at the Z point (see
Fig. 1 in SI), contradicting the experiment measurement. To overcome
this problem, the HSE06 hybrid functional \cite{hse06} is utilized and the resultant band
structure is displayed in Fig. \ref{fig1}(c). It clearly shows that
orthorhombic As is an intrinsic direct bandgap semiconductor with
bandgap of about 0.39 eV, comparable to the experimental value.

At the Z point, one valence band (VB) and
one conduction band (CB) near the Fermi level disperse quite 
strongly along the Z-Q and Z-$\Gamma$ directions (Fig.
\ref{fig1}(c)). As a result, very small effective masses of carriers can
be highly expected.  Based on the nearly free
electron model, the effective mass of carrier can be evaluated as 
$m^{*}=\hbar^{2}/(\partial^{2}E/\partial k^{2})$ with $\hbar$, $E$,
and $k$ being Planck's constant divided by 2$\pi$, the band energy, and the crystal
momentum, respectively. Using the above equation, almost equal effective masses of 0.13 $m_{0}$
(where $m_{0}$ is the free-electron mass) are obtained
 for both electrons and holes along the Z-Q direction, whereas those along the Z-$\Gamma$ direction
are smaller (0.10 $m_{0}$) for electrons and larger
(0.26 $m_{0}$) for holes. Relatively larger values are found along
the direction Z-T${'}$-A$^{'}$, where the carrier effective masses are 1.26
$m_{0}$ and 1.70 $m_{0}$ for electrons and holes, respectively. All
these values are consistent with the experimental report
\cite{cooper}. 
 Thus,  such small effective masses obtained for orthorhombic As would
 set a crucial precondition for few-layer arsenene to exhibit a high
 carrier mobility, as discussed below.

\begin{figure}
\includegraphics[width=16cm]{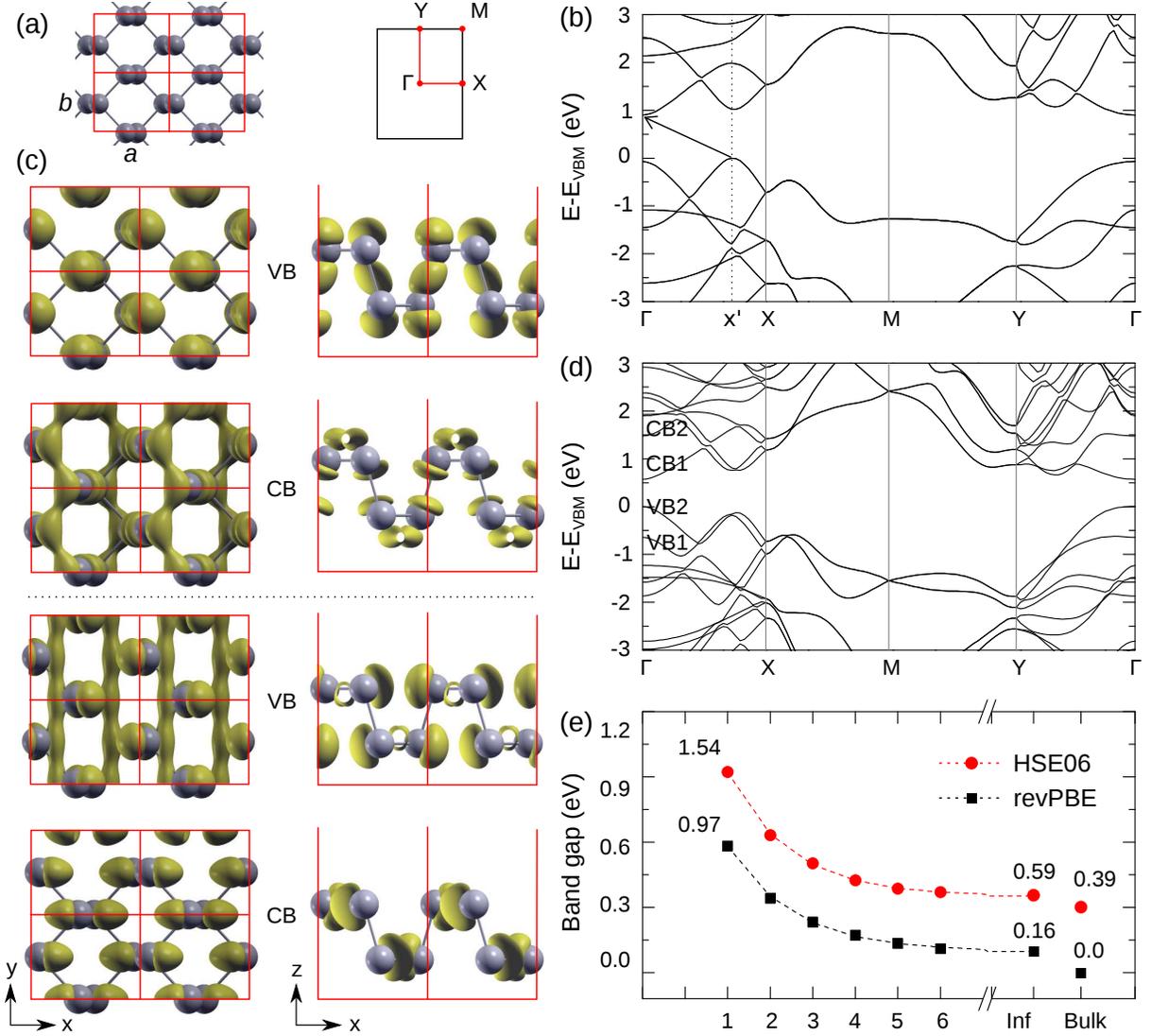}
\caption{(color online) (a) Top view of the monolayer arsenene and the
  associated Brillouin zone. (b) Band structures of monolayer
  arsenene. The arrow shows an indirect band gap.  (c) Spatial
  wavefunctions of VBM and CBM at the $\Gamma$ (top panel) and x$'$
  (labeled in b, bottom panel) points for monolayer arsenene. The
  isovalues are set to 0.05 $e${\AA}$^{-3}$ (d) Band structure of
  bilayer arsenene.  The bands VB1(2) and CB1(2) are labeled as well. 
 (e) The fundamental band gaps as  a function of
  the layer number at the $\Gamma$ point. The bandgap values are
  marked for the monolayer system, for the extrapolation of our
  results and for the real bulk As. It should be noticed that the
  bandgap of monolayer is the direct value at the $\Gamma$ point, not
  that as indicated by the arrow in (b). All results are
  obtained under the  optimized structures with the revPBE-vdW
  functional.}  
\label{fig2}
\end{figure}

In principle, arsenene can be fabricated through
mechanically isolating the orthorhombic bulk counterpart. Here, we
validate it by using first principles calculations, as shown in 
Fig. \ref{fig2}(a). Arsenene is indeed thermally stable since no imaginary
frequency is observed in its phonon dispersion curve (see Fig. 2 in
SI). In addition, we also check its stability from the cohesive
energy. Although the cohesive energy of the grey As is more stable by
about 0.15 eV/atom  than that of 
orthorhombic one, 
the relatively smaller difference  of around 0.01 eV/atom is
obtained for their respective monolayers, indicating the possible
existence of orthorhombic arsenene in experiments. 
The lattice parameters $a$ = 4.80 {\AA} and $b$ =
3.68 {\AA} are obtained for arsenene. Compared with the bulk phase, 
the lattice constant $a$ is elongated by 2.8 percent in arsenene,
while $b$ is shortened by 0.8 percent. As a result, the bond angles
$\theta$1 and $\theta$2 are changed accordingly, i.e., $\theta$1 =
100.92$^{\rm o}$ and $\theta$2 =  94.52$^{\rm o}$, leaving the bond
lengths $R$1 = 2.49 {\AA} slightly decreased
 and $R$2 = 2.51 {\AA} nearly unchanged.
  Such modification in structural parameters stems
from the lack of interlayer interactions when it goes from bulk 
to monolayer. However, the effect on electronic
structure is huge, as discussed below.

The calculated band structure of arsenene is displayed in
Fig. \ref{fig2}. It is interesting to note that arsenene is an indirect
semiconductor with bandgap of $\sim$0.90 eV
 since the valence band maximum (VBM) and
the conduction band minimum (CBM) occur at different
crystal points x$'$ and $\Gamma$. Our results
are consistent with that reported by Kamal and Ezawa \cite{kamal},
except that our gap value is slightly larger by $\sim$
0.05-0.07 eV. This originates from the different structural parameters
obtained from different functionals. 
Such an indirect bandgap in arsenene contrasts
 strikingly with phosphorene
with a direct bandgap. To unveil this  
difference, we carefully check those structural parameters between
them \cite{ji}. Two main changes occur for arsenene: (i) The 
bond lengths $R$1 and $R$2 increase by $\sim$0.2 {\AA} in arsenene, indicating
the relatively weak covalent bonding of As-As compared with P-P; To
still keep the stability, (ii) the bond angles $\theta$1 and $\theta$2 shrink
by 1.48-2.59$^{\rm o}$. We emphasize that this is a natural
consequence as the heavier As atom behaves a weaker covalent characteristic 
compared to P. This is also confirmed in the similar monolayer of
antimony where the structural parameters largely change and the VB at
the x$'$ point significantly exceeds that at the $\Gamma$ point (not
shown for brevity), claiming an obvious indirect bandgap. 
To further understand the underlying physics, we
map out the respective wavefunctions of VB and CB at 
the  x$'$ and $\Gamma$ points, as shown in
Fig. \ref{fig2}(c). Our findings are very insightful. For VBs, the
bonding feature appears between the atoms linked by $R$1 at the
$\Gamma$ point (see the first panel of Fig. \ref{fig2}(c)), while that
at the x$'$ point occurs between the atoms connected by $R$2
(see the third panel of Fig. \ref{fig2}(c)). The mutual competition between
them dictates the VBM. Obviously, in the case of
arsenene, the VB at the x$'$ point forms a weaker covalent bond and thus
occupies the higher energy level. The opposite situation is found for
CBs at the x$'$ and $\Gamma$ points. This is why arsenene is an
indirect semiconductor.

 Based on the above understanding, it is expected to realize
  the bandgap transition from indirect to direct via changing $R$1 and
$R$2 accordingly. Layer stacking is demonstrated to be an effective
  approach in this express. Once any more layer is added, additional bands are
introduced near the Fermi energy level. For example, at the $\Gamma$
point, the bands VB1 and CB2 appear, as shown in Fig. 2(d).
 Due to the
overlap of wavefunctions, VB1 and VB2 (CB1 and CB2) contribute to the bonding
and antibonding states, respectively.  However, the energy separations
between those states are entirely different. 
At the $\Gamma$ point, this value reaches as high as $\sim$1 eV
for the conduction bands and about 0.5 eV for the valence bands, being
obviously much larger than those at the x$'$ point. 
 This is because the $p$ local
orbital directly couples the neighbor layer at the $\Gamma$ point, but
not for the x$'$ point.  Such an overlap of interlayer $p$ local orbitals 
not only makes the bonding length $R$1 shorter, but also importantly 
pushes the bands CB1 and VB2 closer towards each other compared to
those at the x$'$ point.  
Thus, an indirect-direct bandgap transition is obtained  at
bilayer or multilayer, extending the potential applications of
few-layer arsenene.  This also explains why
the orthorhombic bulk As is a direct bandgap semiconductor. Another
effect of layer-stacking reflects in the decrease of bandgap, sharing
the same mechanism proposed in our previous work \cite{xie_jpcl}. 
As the layer increases, the 
fundamental bandgap decreases from 0.97 eV at monolayer to 0.16 eV at
six-layer, as shown in Fig. \ref{fig2}(e). These bandgap values follow the
exponential decay law (see the fitting dashed line). 
The limit bandgap obtained by extrapolation is
around 0.16 eV (marked as Inf in Fig. \ref{fig2}(e)),
 larger than that we obtained for the bulk calculation. 
This is originated from the structural adjustment from
bulk to few-layer, as suggested by the changed structural parameters in
few-layer arsenene (see Tab. I in SI). More accurate bandgaps of
few-layer arsenene are obtained by the HSE06 hybrid functional, as
shown by the red circles and dashed line in Fig. \ref{fig2}(e). 
While arsenene is of
fundamental importance, practical applications involving this material
may require only a small piece of it or a flake, but not in an infinite
size. In this respect, their ribbons in nanometer scale with
well-defined shape may be crucial for device applications. It is found
that several armchair arsenene nanoribbons possess improved and
tunable direct bandgaps 
following the quantum confinement effect \cite {xie_jap} (see Fig. 3 in SI
for more details).

In the following, we restrict our attention to
 carrier mobilities in few-layer arsenene. 
Based on  the so-called acoustic phonon
limited approach \cite{prb_kaasbjerg}, the carrier mobility reads
\be
\mu_{2D}=\frac{e\hbar^{3}C_{2D}}{k_{B}Tm^{*}m_{a}(E^{i}_{l})^{2}},
\ee
where $e$ is the electron charge, $\hbar$ is Plank's constant
divided by 2$\pi$, $k_{B}$ is Boltzmann's constant and $T$ is the
temperature. $m^{*}$ ($m^{*}_{x}$ or $m^{*}_{y}$) is the effective
mass in the transport direction and $m_{a}$ is the averaged effective
mass determined by $m_{a}=\sqrt{m^{*}_{x}m^{*}_{y}}$. $E^{i}_{l}$ is the deformation
potential constant of VBM for hole or CBM for electron along the
transport direction, defined by 
$E^{i}_{l}=\Delta V_{i}/(\Delta l/l_{0})$. Here $\Delta V_{i}$ is the
energy change of the $i^{th}$ band under proper cell compression and
dilatation, $l_{0}$ is the lattice constant in the transport direction
and $\Delta l$ is the deformation of $l_{0}$. The elastic modulus
$C_{2D}$ of the longitudinal strain in the propagation directions
(both $x$ and $y$) of the longitudinal acoustic wave is given
by $(E-E_{0})/S_{0}=C_{2D}(\Delta l/l_{0})^{2}/2$, where $E$ is the
total energy and $S_{0}$ is the lattice volume at equilibrium for a
2D system.

Besides the effective mass, other two properties of the deformation
potential and the 2D elastic modulus are also involved in the calculation of carrier
mobility, as described in Eq. (1). 
First principles
calculations based on the functional of
revPBE-vdW are taken to yield these properties together with the
carrier mobility, as illustrated in Tab. II.
Since the VBM of the monolayer is located at the x$'$ point but not at
the $\Gamma$ point (Fig. 2(b)), we neglect the  discussion of electron mobility as
listed in Tab. II for the monolayer. In the bilayer, the electron
mobility exhibits a value of 0.26-0.40$\times$10$^{3}$
cm$^{2}$V$^{-1}$s$^{-1}$ along the
$\Gamma$-X direction, being about four times higher than that along
the $\Gamma$-Y direction. This triggers a strongly directional
anisotropy, mainly caused by the relatively small effective mass
and deformation potential.
In addition, as the layer further increases, the electron
mobility monotonously increases and reaches 2.17-2.66(0.45-0.51)$\times$10$^{3}$
cm$^{2}$V$^{-1}$s$^{-1}$ along the $\Gamma$-X(Y) direction at
six-layer. The directional anisotropy remains nearly unchanged.
In the case of hole, in monolayer, the hole mobility of the $\Gamma$-Y
direction is slightly larger than that of
the $\Gamma$-X direction. But in multilayer, the $\Gamma$-X direction holds
much higher hole mobilities, reaching as high as
$\sim$4$\times$10$^{3}$ cm$^{2}$V$^{-1}$s$^{-1}$ at five-layer. This
is a huge value if one remembers the typical carrier mobilities of
200-500 cm$^{2}$V$^{-1}$s$^{-1}$ in MoS$_{2}$ \cite{fivaz,radis}. Such
a high hole mobility  in few-layer arsenene benefits from the smaller
deformation potential. The deformation
potential reaches as small as $\sim$1.7 eV at four-layer, which is about two times
smaller than the value of 3.9 eV in MoS$_{2}$ \cite{kaasbjerg}.
It is also noticed that around 1-2 orders of magnitude smaller in hole
mobilities are obtained along the $\Gamma$-Y direction, resulting in a
large directional anisotropy of hole mobility. It is well known that
the higher carrier mobility plays a crucial role in the performance of
devices. Thus, few-layer arsenene with proper bandgaps and high
carrier mobilities are potential for practical applications in semiconducting
industry.

In conclusion, we have investigated the electronic properties of few-layer
arsenene from first principles calculations. Our results show that
few-layer arsenene possesses carrier mobilities as high as several
thousand square centimeters per volt-second, which exhibits a high
anisotropy.   Combining such superior carrier
mobilities with the tunable bandgaps around 1 eV, few-layer
arsenene is an ideal semiconducting material for device applications
in semiconducting industry.

We thank Dr. Wei Ji (RUC) for valuable communications. 
This work was supported by the National Basic Research Program of
China under Grant No. 2012CB933101 and the National Science
Foundation under Grant No. 51372107, No. 11104122 and No. 51202099.
 We also acknowledge this work as done on Lanzhou
University's high-performance computer Fermi.

$^{*}$ sims@lzu.edu.cn

$^{\dagger}$ xueds@lzu.edu.cn

\clearpage

\clearpage

\clearpage

\begin{table*}[!hbp]
\caption{The optimized lattice constants ($a$, $b$, and $c$) and structural
  parameters ($R$1,$R$2, $\theta$1, and $\theta$2) for the
  orthorhombic bulk As  under five functionals. The corresponding
  values in experiment are given as well.}
\begin{tabular}{cccccccc}
\hline
\hline
Functional & $a$ ({\AA}) & $b$ ({\AA}) & $c$ ({\AA}) & $R$1 ({\AA}) & $R$2 ({\AA})
& $\theta$1 ($^{\rm o}$) & $\theta$2 ($^{\rm o}$) \\
\hline
Expt. \cite{smith} & 4.47 & 3.65 & 11.00 & 2.48 & 2.49 & 98.5 & 94.1 \\
PBE  & 4.67  &3.71  & 11.45  &2.51  &2.52  & 99.89  & 95.08 \\
revPBE-vdW &4.67  & 3.71 & 11.07 & 2.50 & 2.51 & 100.01 & 95.29 \\
optPBE-vdW  & 4.38  &3.72  &10.91  & 2.49  & 2.50 &98.03  &96.30  \\
optB88-vdW  & 4.26  &3.73  & 10.88  & 2.48 & 2.49 & 97.29   & 96.83 \\
optB86b-vdW  & 4.08  &3.74  & 10.79 &2.47  & 2.50 & 95.90 &96.86  \\
\hline
%\hline
\end{tabular}
\label{tab1}
\end{table*}

\clearpage

\begin{table*}[!hbp]
\caption{Carrier mobilities in few-layer arsenene. Types ``e'' and
  ``h'' denote the ``electron'' and ``hole'',
  respectively. $m^{*}_{x}$ ($m^{*}_{y}$) (in unit of $m_{0}$ with
  $m_{0}$ being the static electron mass) represents the effective
  mass along the $\Gamma$-X (Y) direction. $E_{1x}$ ($E_{1y}$) (in unit
  of eV) is the deformation potential at $\Gamma$ point along the
  $\Gamma$-X (Y) direction. $C_{x_{-}2D}$ ($C_{y_{-}2D}$) is the 2D
  elastic modulus for the  $\Gamma$-X (Y) direction which is in unit of
  Jm$^{-2}$. Carrier mobility $\mu_{x_{-}2D}$ ($\mu_{y_{-}2D}$) (in
  unit of 10$^{3}$ cm$^{2}$V$^{-1}$s$^{-1}$) along the  $\Gamma$-X (Y)
  direction is calulated by using Eq. (1) together with $T$ = 300 K.}

\begin{tabular}{cccccccccc}
\hline
\hline
Type & N$_{\rm L}$ & $m_{x}^{*}$/$m_{0}$ &  $m_{y}^{*}$/$m_{0}$ &
$E_{1x}$ & $E_{1y}$ & $C_{x_{-}2D}$& $C_{y_{-}2D}$ &$\mu_{x_{-}2D}$ & $\mu_{y_{-}2D}$ \\
\hline
 e    & 1 &0.23 &1.22 &0.81$\pm$0.17 &3.74$\pm$0.04 &28.99 &74.69 &5.29-12.32&0.17-0.18\\
      & 2 &0.25 &1.38 &5.56$\pm$0.68 &6.75$\pm$0.56 &56.74 &147.35&0.26-0.40&0.07-0.11\\
      & 3 &0.23 &1.39 &3.67$\pm$0.21 &5.21$\pm$0.12 &83.98&218.55 &0.92-1.16 &0.21-0.23\\
      & 4 &0.22 &1.40 &4.24$\pm$0.20 &5.24$\pm$0.05 &109.5&266.64 &0.97-1.18&0.26-0.27 \\
      & 5 &0.19 &1.41 &5.13$\pm$0.17 &5.38$\pm$0.11 &136.90&362.37&1.06-1.21&0.35-0.38 \\
      & 6 &0.18 &1.41 &4.17$\pm$0.11 &5.24$\pm$0.17 &164.61&436.75&2.17-2.66&0.45-0.51 \\
 \\
 h    &1  &0.19 &1.77 &4.06$\pm$0.05 &1.88$\pm$0.05 &28.99 &74.69&0.33-0.35&0.42-0.47 \\
      &2  &0.21 &4.49 &1.96$\pm$0.24 &2.01$\pm$0.06 &56.74 &147.35&1.13-1.89&0.13-0.14\\
      &3  &0.20 &7.54 &2.71$\pm$0.09 &5.15$\pm$0.09 &83.98 &218.55&0.93-1.08&0.018-0.019\\
      &4  &0.20 &10.08&2.43$\pm$0.04 &5.53$\pm$0.05 &109.53&266.64&1.35-1.44&0.012-0.013 \\
      &5  &0.18 &11.44&1.70$\pm$0.05&5.79$\pm$0.04&136.90&362.37&3.68-4.19&0.013-0.014\\
      &6  &0.16 &11.44&3.09$\pm$0.21 &6.36$\pm$0.11&164.61&436.75&1.49-1.94&0.014-0.015\\
\hline
%\hline
\end{tabular}
\label{tab2}
\end{table*}


\begin{thebibliography}{99}

\bibitem{nl_cui} Y. Cui, Z. Zhong, D. Wang, W. U. Wang, and
  C. M. Lieber, \nl {\bf 3}, 149 (2003).

\bibitem{sci_nov} K. S. Novoselov, A. K. Geim, S. V. Morozov,
  D. Jiang, Y. Zhang, S. V. Dubonos, I. V. Grigorieva, and  A. A. Firsov,
  \sci {\bf 306}, 666 (2004).

\bibitem{li_nn} L. Li, Y. Yu, G. J. Ye, Q. Ge, X. Ou, H. Wu, D. Feng,
  X. H. Chen, and Y. Zhang,   \nnt {\bf 9}, 372 (2014).

\bibitem{ji} J. Qiao, X. Kong, Z.-X. Hu, F. Yang, and W. Ji, \nc
  {\bf 5}, 4475 (2014).

\bibitem{nl_liang} L. Liang, J. Wang, W. Lin, B. G. Sumpter,
  V. Meunier, and M. Pan, \nl  {\bf  14}, 6400 (2014).

\bibitem{jamieson} J. C. Jamieson, \sci {\bf   139},
  1291 (1963). 

\bibitem{wittig} J. Wittig and B. T. Matthias, \sci {\bf 160},
  994 (1968).

\bibitem{liu_acs} H. Liu, A. T. Neal, Z. Zhu, Z. Luo, X. Xu,
  D. Tom\'aek, and P. E. Ye,  \acsnn {\bf 8}, 4033 (2014).

\bibitem{nl_fei0} R. Fei, A. Faghaninia, R. Soklaski, J.-A. Yan,
  C. Lo, and L. Yang, \nl {\bf  14}, 6393 (2014).

\bibitem{jiang} J.-W. Jiang and H. S. Park, \nc {\bf 5},
  4727 (2014). 

\bibitem{nl_fei} R. Fei and L. Yang, \nl {\bf 14}, 2884 (2014).

\bibitem{kamal} C. Kamal and M. Ezawa, arXiv:1410.5166v1.

\bibitem{jianw} J. Han, J. Xie, Z. Zhang, D. Yang M. Si, and D. Xue,
  to be published in Appl. Phys. Exp.

\bibitem{takeda} K. Takeda and K. Shiraishi, \prb {\bf  50}, 14916 (1994).

\bibitem{prl_cahangirov} S. Cahangirov, M. Topsakal, E. Akt\"urk,
  H. Sahin, and S. Ciraci, \prl {\bf 102}, 236804 (2009).


\bibitem{prl_vogt} P. Vogt, P. D. Padova, C. Quaresima, J. Avila,
  E. Frantzeskakis, M. C. Asensio, A. Resta, B. Ealet, and G. L. Lay,
  \prl {\bf 108}, 155501 (2012).

\bibitem{prl_fleurence} A. Fleurence, R. Friedlein, T. Ozaki,
  H. Kawai, Y. Wang, and Y. Yamada-Takamura, \prl {\bf 108},
  245501 (2012).

\bibitem{xu} J. H. Xu, E. G. Wang, C. S. Ting, and W. P. Su,
  \prb {\bf 48}, 17271 (1993).

\bibitem{krebs} H. Krebs, W. Holz, and K. H. Worms, Chem. Ber.
  {\bf  90}, 1031 (1957).

\bibitem{greaves} G. N. Greaves, S. R. Elliott, and E. A. Davis,
  Advcances in Physics {\bf 28}, 49 (1979).


\bibitem{pollard} W. B. Pollard and J. D. Joannopoulos, \prb {\bf 19},
  4217 (1979).

\bibitem{si} See Supporting Information at http://iopscience.iop.org.

\bibitem{smith} P. M. Smith, A. J. Leadbetter, and A. J. Apling,
 Philos. Mag. B {\bf 31}, 57 (1975).

\bibitem{rev} M. Dion, H. Rydberg, E. Schr\"oder, D. C. Langreth, and
  B. I. Lundqvist, \prl {\bf 92}, 246402 (2004).

\bibitem{pbe} J. P. Perdew, K. Burke, and M. Ernzerhof, \prl
  {\bf 77},  3865 (1996).

\bibitem{hse06} J. Heyd, G. E. Scuseria, and M. Ernzerhof,  \jcp
  {\bf    118}, 8207 (2003).

\bibitem{cooper} O. S. Cooper and A. W. Lawson,  \prb {\bf
  4}, 3261 (1971).

\bibitem{xie_jpcl} J. Xie, Z. Y. Zhang, D. Z. Yang, D. S. Xue,
  and M. S. Si, \jpcl {\bf 5}, 4073 (2014).

\bibitem{xie_jap} J. Xie, M. S. Si, D. Z. Yang, Z. Y. Zhang, and
  D. S. Xue, \jap {\bf 116}, 073704 (2014).

\bibitem{prb_kaasbjerg} K. Kaasbjerg, K. S. Thygesen, and A.-P. Jauho,
 \prb {\bf 87}, 235312 (2013).

 \bibitem{fivaz} R. Fivaz and E. Mooser, \pr {\bf 163},
  743 (1967). 

\bibitem{radis} B. Radisavljevic, A. Radenovic, J. Brivio,
  V. Giacometti, and A. Kis,  \nn {\bf 6}, 147 (2011).

\bibitem{kaasbjerg} K. Kaasbjerg, K. S. Thygesen, and K. W. Jacobsen,
 \prb {\bf 85}, 115317 (2012).
\end{thebibliography}
\end{document}